\documentclass[aps,prb,amsmath,amssymb,twocolumn,notitlepage]{revtex4-2}

\usepackage{times}
\usepackage[T1]{fontenc}
\usepackage{bm}
\usepackage{bbm}
\usepackage{float}
\usepackage{mathrsfs}
\usepackage[utf8]{inputenc}
\usepackage{pdfsync}
\usepackage[colorlinks=true,bookmarks=true,pdfborder={0 0 0},linkcolor=blue,urlcolor=blue,breaklinks=true,bookmarksnumbered=true]{hyperref}

\def \hs{\hat{s}}

\def \hu{\hat{u}}

\def \hO{\hat{O}}

\def \hrho{\hat{\rho}}

\def \mH{\mathcal{H}}

\def \mL{\mathcal{L}}

\def \pt{\partial}
\def \si{\sigma}

\def \tr{\text{Tr}}
\def \Xs0{X^{\sigma 0}}
\def \X0s{X^{0 \sigma}}

\newcommand{\abs}[1]{\lvert#1\rvert}

\newcommand{\ev}[1]{\mbox{$\langle #1 \rangle$}}

\newcommand{\ket}[1]{\mbox{$| #1 \rangle$}}

\newcommand{\ua}{\uparrow}
\newcommand{\da}{\downarrow}

\begin{abstract} Variational methods are of fundamental importance and widely used in theoretical physics, especially for strongly interacting systems. In this work, we present a set of variational equations of state (VES) for pure states of an interacting quantum Hamiltonian. The VES can be expressed in terms of the variation of the density operators or static correlation functions. We derive the algebraic relationship between a known pure state density matrix and its variation, and obtain the VES by applying this relation to the averaged Heisenberg-equations-of-motion for the exact density matrix. Additionally, we provide a direct expression of the VES in terms of correlation functions to make it computable. We present three nontrivial applications of the VES: a perturbation calculation of correlation functions of the transverse field Ising model in arbitrary spatial dimensions, a study of a longitudinal field perturbation to the one-dimensional transverse field Ising model at the critical point and variational calculation of magnetization and ground state energy of the two-dimensional spin-1/2 Heisenberg model on a square lattice. For the second one, our results not only recover the scaling limit, but also indicate the possibility of continuous tuning of the critical exponents by adjusting the longitudinal fields differently from the scaling limit. For the Heisenberg model, we obtained results numerically comparable to established results with simple calculations. The VES approach provides a powerful and versatile tool for studying interacting quantum systems. \end{abstract}   %revtex abstract
\date{\today}
\title{Variational Equations-of-States for Interacting Quantum Hamiltonians}
\hypersetup{
 pdfauthor={Wenxin Ding\(^{1*}\)},
 pdftitle={Variation Equations of States for Interacting Quantum Hamiltonians},
 pdfkeywords={},
 pdfsubject={},
 pdfcreator={wxding@ahu.edu.cn},
 pdflang={English}}
\begin{document}

\title{Variational Equations-of-States for Interacting Quantum Hamiltonians} \author{Wenxin Ding\(^{1*}\)}
 \date{\today} \affiliation{$^1$School of Physics and Optoelectronic Engineering, Anhui University, Hefei, Anhui Province, 230601, China}
  \email{wxding@ahu.edu.cn}
 \maketitle
%\begin{abstract} test \end{abstract}

\emph{Introduction.}---Variational methods are of great importance to theoretical physics. Numerous fundamental physical laws such as the principle of least action are variational in nature. It has been instrumental in deriving fundamental equations of motion, such as the Euler-Lagrange equations in Lagrangian mechanics. Practically, variational methods are at the core of many optimization problems. In quantum mechanical problems, the variational Lippmann-Schwinger equation (LSE)\cite{lippmann-1950-variat-princ} is commonly used to solve scattering-related problems across different fields of physics. For interacting problems, many standard methods, such as Hartree-Fock methods\cite{hartree-1928-wave-mechan}, are fundamentally variational. For strongly interacting many-body systems, variational methods are particularly indispensable, as both perturbation methods and exact numerical solutions are very limited.
For example, for problems as simple as the Kondo problem\cite{Kondo1964,Hewson1997}, a single magnetic impurity in metals, the complete solution requires complex machinery as the Wilson's numerical renormalization group (NRG) method\cite{wilson-1975-renor-group}, which are understood as variational in nature nowadays\cite{pizorn-2012-variat-numer}. Methods inspired by NRG, such as the density matrix renormalization group (DMRG)\cite{white-1992-densit-matrix,scholl-dmrg-2005}, matrix product state (MPS)\cite{verstraete-2007-mps}, tensor-network states \cite{cirac-2009-renor-tensor}, etc., all fall into the category of variational methods.

While these methods greatly improved our understanding of many strongly correlated systems, some categories of problems still remain challenging for various reasons. The typical difficulty is the combination of higher dimensions (D>1) and gaplessness or criticality. For example, the nature of the \(J_1-J_2\) Heisenberg models on the square lattice near its putative critical point or region, \(J_2/J_1 \sim 1/2\)\cite{chandra-1988-possib-spin}, is still under heated debate. The strong quantum entanglement brought by such combination greatly increases the computational costs. To tackle such problems, new methods and ideas are needed. Algebraic approaches to quantum statistical mechanics\cite[][]{book-emch-1972-algebraic-methods,book-bratteli-1997-operat-algeb} provide a powerful framework for such purposes.

In this letter, we employ a \emph{complete  operator basis set} (COBS) to represent the density matrices, which was first introduced by J. Schwinger\cite{schwinger-1960-unitar-operat-bases} and U. Fano\cite{fano-1957-descr-states}, and use operator algebras to derive a set of variational equations of states (VES) for arbitrary interacting quantum Hamiltonians on lattices. In COBS representation, a density matrix is specified directly by the many-body correlation functions, e.g. expectation values of elements of COBS. Then the VES are a set of algebraically coupled equations for variations of these correlation functions. Such VES is a generalization of the variational LSE for interacting many-body Hamiltonians. It is also a formal solution to the quantum version of time-independent BBGKY hierarchy equations\cite{fesciyan-1973-bbgky-hierar,gallavotti-1968-on-mechanical}, akin that the
LSE is a formal solution to Schr\"odinger's equation of a scattering potential.
Moreover, it is expressed directly in terms of correlation functions instead of wavefunction coefficients, which make it more amenable for further optimization, such as energy minimization.

%The rest of the work is organized as the following. We first introduce an algebraic approach to study variations of density matrices. Then we cast such algebraic equations of density operators into a component form, i.e. in terms of correlation functions and their variations. After that, we discuss both the perturbative and variational schemes within this framework. We subsequently apply the VES to three representative problems as benchmarks: i) the transverse field Ising model, ii) the transverse field Ising model at critical point subjecting to a longitudinal field perturbation and iii) the \(J_1\) spin-\(1/2\) Heisenberg model on the square lattice. In all three cases, we obtain results in quantitative agreement with established ones with simple calculations. Finally, we discuss the potential of application of VES to unsolved problems.

\emph{Algebra of Density Operators}.---A COBS\cite{schwinger-1960-unitar-operat-bases} is a set of operators \(\mathscr{U} = \{ \hu^\alpha \}\) satisfy both a orthogonality condition and a completeness condition.
The orthogonality condition requires that \(\tr ((\hu^\alpha)^\dagger \hu^\beta) = C ~\delta_{\alpha \beta}\), where \(C\) is a normalization factor which will be taken as 1 for convenience unless noted otherwise. The completeness requires that the product of elements from COBS satisfy a set of \emph{closed} algebraic relations \(\hu^\alpha \hu^\beta = \sum_{\gamma} a^{\alpha \beta}_\gamma \hu^{\gamma}\), where \(a_{\alpha \beta}^\gamma \in \mathbb{C}\). Conventionally, it is convenient to decompose the algebras into a symmetric (bosonic) sector and an anti-symmetric (fermionic) sector \([\hu^\alpha, \hu^\beta ] = \sum_{\gamma} b^{\alpha \beta}_\gamma \hu^\gamma, ~\{ \hu^\alpha, \hu^\beta \} = \sum_{\gamma} f^{\alpha \beta}_\gamma \hu^\gamma\).

The elements of COBS generally can be non-Hermitian. In this work, we always assume a Hermitian COBS. For example, a single quantum spin\(-1/2\) can described by \(\mathscr{U} = \{\si_0, ~\si^x, ~ \si^y, ~\si^z \}\) where \(\si^\alpha\)'s are Pauli matrices. For quantum many-body systems on lattices, the many-body COBS can be constructed by direct products of the local COBS \(\mathscr{U} = \otimes_i \mathscr{U}_i = \{ \hu^\alpha_i, ~ \dots, ~\hu^\alpha_i \hu^\beta_j, ~ \dots \}\).

With both the complete and the orthogonal conditions, any operator \(\hO\) of \(\mathscr{H}\) can be expanded into a sum of \(\hu^\alpha\)s, \(\hO = \sum_\alpha \tr (\hO \hu^\alpha) \hu^\alpha\).
Therefore, any density operators can be expressed in a component form via the correlation functions
\begin{align}
\hat{\rho} = \sum_\alpha \tr (\hat{\rho} \hu^\alpha) \hu^\alpha = \sum_\alpha \ev{\hu^{\alpha}} \hu^\alpha
\label{eq:1}.
\end{align}

While Eq. \eqref{eq:1} is valid for arbitrary density operators \(\hat{\rho}\)s, we restrict to pure states within this work which imposes further constraints on \(\hat{\rho}\)s.
For eigenstates of a given Hamiltonian \(\mH\) with an eigen energy \(E\), the corresponding density operators \(\hat{\rho}\) satisfy the following algebraic equations: i) the equilibrium state condition \([\hat{\rho}, \mH] = 0\), ii) the pure state condition \(\hat{\rho}^2 = \hat{\rho}\) and iii) eigenstate condition \(\{\hat{\rho},\mH\} = 2 E \hat{\rho}\).
These constraints can be turned into constraint equations for correlation functions, i.e. the component form, by multiplying the equations by a \(\hu^{\alpha}\) and taking the trace: \(i\pt_t \ev{\hu^{\alpha}} = 0 = \ev{[H, \hu^{\alpha}]}, \ev{\hu^{\alpha} H} = E \ev{\hu^{\alpha}}\).

While we shall use condition iii) to compute energy perturbations, which is equivalent to using \(E = \tr \left[ (\hrho_0 + \delta \hrho) (\mH_0 + g \mH') \right]\), it contains more information which can be used to minimize the energy variation. However, within this work, we assume a Lippmann-Schwinger type scenario and without systematic discussion of energy minimization. Further energetic variations will be discussed in future works.

\emph{Perturbative Equations of States}.---
First, we discuss a perturbation theory using the algebraic equations of \(\hrho\) for an eigenstate of \(\mH = H_0 + g \mH'\), where \(g\) is a dimensionless parameter.
We assume the following perturbative expansions are valid: \(\hrho = \hrho_0 + \delta \hrho = \hrho_0 + g \hrho_1 + \frac{g^2}{2} \hrho_2 + \dots = \sum_{n=0}^\infty \frac{g^n}{n!} \hrho_n\), \(E = E_0 + \delta E = E_0 + g E_1 + \frac{g^2}{2} E_2 + \dots =  \sum_{n=0}^\infty \frac{g^n}{n!} E_n\).
Keeping terms up to \(\mathcal{O}(g^2)\), we have the following perturbation equations. At  \(\mathcal{O}(g^1)\), we obtain
\begin{align}
[\hrho_1, \mH_0] + [\hrho_0, \mH'] =  0, \quad \{\hrho_0, \hrho_1 \} = \hrho_1. \label{eq:2}
\end{align}

To make the method computable, we need to turn the operator equations into equations of states, or correlation functions, by making use of COBS representation. The components of the perturbed density matrix shall be computed from the unperturbed ones. Writing \(\hrho_1 = \sum_{\hu^{\alpha}} \ev{\hu^{\alpha}}_1 \hu^{\alpha}\), multiplying the first equation of Eq. \eqref{eq:2} by \(\hu^{\alpha_0}\), taking the trace and inserting the second one, we find a perturbation equation in terms of the unperturbed correlation functions
\begin{align}
\ev{[\mH',\hu^{\alpha_0}]}_0 + \sum_\alpha  \ev{\{[\mH_0,\hu^{\alpha_0}], \hu^{\alpha}\}}_0 \ev{\hu^{\alpha}}_1 = 0. \label{eq:3}
\end{align}
Eq. \eqref{eq:9} can be represented in a matrix form as
\begin{align}
 [[\mathcal{L}_0]] \cdot [\rho_1] + [V]_{0} = 0,\label{eq:4}
%\text{where} \nonumbering
\end{align}
with \([\rho_1] = \{\ev{\hu^{\alpha_1}}_1, \dots, \ev{\hu^{\alpha_i}}_1, \dots\}^T\), \([V]_{0} = \{\ev{[\mH',\hu^{\alpha_1}]}_0, \dots, \ev{[\mH',\hu^{\alpha_i}]}_0, \dots \}^T\) and \([[\mathcal{L}_0] ]_{\alpha \alpha_0} =  \ev{\{[\mH_0,\hu^{\alpha_0}], \hu^{\alpha}\}}_0\). Therefore, we can formally solve \(\hrho_1\) exactly by inverting the matrix \([[\mathcal{L}_0] ]\).

We can interpret Eq. \eqref{eq:4} as follows. In Eq. \eqref{eq:2}, the first term is the time-evolution of \(\hrho_1\) under the action of \(\mH_0\). Consider the Liouville-von Neumann equation for \(\hrho_1\): \(i \hbar \frac{d \hrho_1(t)}{dt} = [\mH_0,\hrho_1(t)] \Rightarrow \rho_1(t) = \exp[-i \mL_0 t / \hbar] ~ \hrho_1(0)\). Then taking \(\text{$t \rightarrow 0$ gives } [\mH_0,\hrho_1] = \mL_0 \hrho_1\). The action of \(\mathcal{L}_0\) thus should be identified as a representation of the Liouvillian superoperator due to \(\mH_0\) on \(\hrho_1\).

\emph{Variational Equations of States}.---
Following the construction of LSE, it is natural to promote the perturbation theory to a variational one. We still consider Hamiltonians of the same form, \(\mH = \mH_0 + g\mH'\) with \(\mH_0\) has a known ground state \(\ket{\Phi_0}\) specified by \(\ev{\hu^\alpha}_0\). However, \(\mH_0,~\mH,~E_0,~E,~\hrho_0,~\hrho\) satisfy the following \emph{exact equations} instead of perturbative ones (subscript is ignored): \([\hrho,\mH] = 0,~\{\hrho, \mH\} =2E \hrho,~ \hrho = \hrho_0 + \delta\hrho, \quad E = E_0 + \delta E\). With some algebra, we obtain the following exact recursion relation for \(\delta \hrho\):
\begin{align}
\{\hrho_0, \delta \hrho\} = \delta \hrho - \delta \hrho^2. \label{eq:5}
\end{align}
The key difference with 1st order perturbation is the relation \(\{\hrho_0, \delta \hrho\} = \delta \hrho - \delta \hrho^2\). The first term of RHS simply yields the same contribution to \(\delta \hrho\) as the first order perturbation. Explicitly, we have
\begin{align}
  \begin{split}
\ev{[\mH',\hu^{\alpha_0}]}_0 + \sum_\alpha  \ev{\{[\mH,\hu^{\alpha_0}], \hu^{\alpha}\}}_0 \ev{\hu^{\alpha}}_1 = 0 \Leftrightarrow \\
[V]_{0,\alpha_0} +  ([[\mL]] \cdot [\delta \hrho])_{\alpha_0} + \tr \left[  \hu^{\alpha_0}  [\delta \hrho^2, \mH] \right] = 0.
  \end{split}
  \label{eq:6}
\end{align}
The key difference here comparing to the perturbation theory is that now \(\mathcal{L}\) is generated by the full Hamiltonian instead of the unperturbed part \(\mH_0\).

To treat the last term, we apply Eq. \eqref{eq:5} iteratively.
\begin{align}
  \begin{split}
  & \tr \left[  \hu^{\alpha_0}  [\delta \hrho^2, \mH] \right]
  %& = \tr \left[  \hu^{\alpha_0} \{ \delta \hrho, [\{\hrho_0, \delta \hrho\}, \mH]\} \right] + \tr \left[\hu^{\alpha_0} [\delta \hrho^3, \mH] \right]\\
   = I^{\hu^{\alpha_0}}_{2} + \tr \left[  \hu^{\alpha_0}  [\delta \hrho^3, \mH] \right]\\
   & = \sum_{n=2}^{N} I^{\hu^{\alpha_0}}_{n} + \tr \left[  \hu^{\alpha_0}  [\delta \hrho^{N+1}, \mH] \right] = \dots_{N \rightarrow \infty},
  %& = \sum_{n=2}^N I_{n} + \ev{\delta \hu^{\alpha_0}}_{N+1}. %\ev{\delta \hu^{\alpha_0}}_2 =
  \end{split}\label{eq:7}
\end{align}
where \(I_{n}\) reads
\begin{align}
I^{\hu^{\alpha_0}}_n =  \ev{ \left\{ \delta \hrho  , [\mH, \{ \dots \{\hu^{\alpha_0}, \delta \hrho \} \dots \} , \delta \hrho \}]\right\} }_0.\label{eq:8}
\end{align}
Note that the first order term is consistent and can be denoted as \(I^{\hu^{\alpha_0}}_1\).

Similarly as the order \(\mathcal{O}(1)\) perturbation, all \(I^{\hu^{\alpha_0}}_n\) can be considered as the application of \(\delta \hrho^n\)'s Liouville operators due to the full Hamiltonian \(\mH\): \([\mH, \delta \hrho^n] = \mL_n \delta \hrho^n, ~ I_n = [[\mL] ] \cdot [ [ \delta \hrho ^n ] ]\).

Therefore, the nonlinear term \(\delta \hrho^2\) can be expressed iteratively as integration with correlation functions of the unperturbed state. If we can take \(N \rightarrow \infty\), Eq. \eqref{eq:6} becomes exact:
\begin{align}
 [V]_{0,\alpha_0} +  \sum_{n=1}^{\infty} ([[\mL_n]] \cdot [\delta \hrho^n])_{\alpha_0} = 0.\label{eq:9}
\end{align}
which can be considered as an analogue of LSE for interacting many-body Hamiltonians.

\emph{A Linked Cluster Theorem for VES}.---
In the component form of VES, there is clearly a lot of redundant equations. To simplify the methods, we need the eliminate the irrelevant ones. For example, consider the second order part of VES, \(\{\hu^\alpha, [\mH, {\hu^{\alpha_0}, \hu^\beta}]\}\). Observe that, for bosonic operators, when \(\hu^{\alpha_0}\) and \(\hu^\beta\) do not overlap, we simply get \(\hu^{\alpha_0}\times \hu^\beta\) out of the anti-commutator. If \(\hu^{\alpha_0}\) survives the commutator with \(\mH\), we always have at least one nonzero contributing term to the VES (\(\propto \delta\ev{\hu^\alpha} \delta \ev{\hu^\alpha }\)) for \(\hu^\alpha = \hu^\beta\).

To obtain a analogue of the linked-cluster theorem for VES, we simply consider \(\hu^{\alpha_0} = 1\), which leads to the following identity
\begin{align}
  \begin{split}
    \ev{[\delta H, 1]}_0 = 0 = - \sum_{n} I^0_n, \\
    I^0_n = \ev{ \{\delta \hrho, [\mH, \{ \dots \{1, \delta \hrho\}, \dots, \delta \hrho \} ] \}}_0.
  \end{split}
\label{eq:10}
\end{align}
We propose that subtracting Eq. \eqref{eq:10} from the VES yields the connected VES, which requires that \(\hu^\beta\) to overlap (sharing at least one lattice site) with the chosen \(\hu^{\alpha_0}\). We should note that this linked cluster theorem does \emph{not} apply to the 1st order VES.

\emph{Cumulant COBS}.---
To further simplify the equations, we construct a cumulant COBS when a state \(\rho_0\) is given. Consider \(\rho_0\) represented by the expectation values of a direct product COBS \(\{ \ev{\hu^\alpha_i}_0, \dots, \ev{\hu^\alpha_i \hu^\beta_j}_0, \dots \}\). For most quantum states, \(\ev{\hu^\alpha_i \hu^\beta_j}_0 - \ev{\hu^\alpha_i}_0 \ev{\hu^\beta_j}_0 = \ev{\hu^\alpha_i \hu^\beta_j}_c \neq = 0\). However, it is inconvenient in such COBS since we cannot easily distinguish between a simple product state, for which \(\ev{\hu^\alpha_i \hu^\beta_j}_c = 0\), from a generic one.

To simplify our VES, we introduce a new COBS with respect to a given \(\rho_0\) as
\begin{align}
(\hu^\alpha_i \hu^\beta_j)_c = \hu^\alpha_i \hu^\beta_j - (\lambda^{\alpha,\beta}_{i,j})^2 \hu^\alpha_i - (1 - (\lambda^{\alpha,\beta}_{i,j})^2) \hu^\beta_j,\label{eq:11}
\end{align}
with \((\lambda^{\alpha,\beta}_{i,j})^2 = \frac{\ev{\hu^\alpha_i}_0^2}{\ev{\hu^\alpha_i}_0^2 + \ev{\hu^\beta_j}_0^2}\). It is straight forward to verify that \(\ev{(\hu^\alpha_i \hu^\beta_j)_c}_0 = \ev{\hu^\alpha_i \hu^\beta_j}_c\).

\emph{Preliminary applications.}-- Next, we apply the VES approach to three important examples as a demonstration of both its validity and high efficiency.

\emph{Application I: perturbation theory 1D transverse field Ising models}.---
Consider the transverse field Ising models on lattices:\(\mH_{\text{TFIM}} = - J \sum_{ij} \hs^x_i \hs^x_j - h_z \sum_i \hs^z_i.\)
First, we do perform perturbation calculations and compare with results of correlation functions in Ref. \cite{pfeuty-1970-one-dimen}. (See SI for details of the calculations). In the \(g = J/h_z \rightarrow 0\) limit, \(\mH_0 = - h_z \sum_i \hs^z_i\), \(\delta \mH = - J \sum_{i} \hs^x_i \hs^x_{i+1}\). The ground state \(\ket{\Phi_0} = \ket{\ua \ua \dots \ua}\).

We find that there is no first order correction to any components of the magnetization: \(\ev{\hs^\alpha}_1 = 0\). This is consistent with the exact results. Expansion of \(\ev{\hs^z}\) in powers of \(J/(2 h^z)\) yields , to the lowest nonzero term, \(\ev{\hs^z} \simeq 1/2 - \lambda^2/8\) with \(\lambda = J/ 2 h^z\), hence \(\ev{\hs^z}_1 = 0, ~ \ev{\hs^z}_2 = - \lambda^2/8\).
The major 1st order correction \(\ev{\hs^x_{i_0} \hs^x_{i_0+1}}_1 = J/(16 h^z) = \lambda/8\) and \(\ev{\hs^y_0 \hs^y_1}_1 = - \lambda / 8\). In the large-J limit, we can also study the large-J limit perturbatively: \(\ev{\hs^z} \simeq \lambda^{-1}/8\).

\emph{Applications II: critical 1D transverse field Ising model perturbed by a longitudinal field}.---
To give a more nontrivial example, we consider the problem of a critical 1d transverse field Ising model subject to perturbation from a small longitudinal field. We consider \(\mH_0 = \mH_{\text{TFIM}} = - \sum_{i} \hs_i^z \hs_{i+1}^z  - g \sum_{i} \hs^x_i\) and \(\delta \mH = - h \sum_{i} \hs^z_i\). While the general problem on the \((g, h)\) plane is considered unsolvable, its integrability is preserved along a special path on the \((g, h)\) plane which is the so-called scaling specified by fixing \(h/\abs{g- g_c}^{-15/8} = const = \delta h\). While this result can be obtained by either dimensional analysis in CFT and perturbative expansion of the partition function\cite{mccoy-1978-two-dimen}, here we provide a new way of obtaining this scaling limit via our generic VES.

In order to preserve the scaling of the Ising universality class, i.e. \(\ev{\hs^z} \sim |1-g/g_c|^{1/8}\) for \(g>g_c\) etc., we let \(h \sim |1-g/g_c|^{a}\).
Working out the VES for \(\hu^{\alpha_0} = \hs^y_{x_0}\), we obtain \(0 = - i h \ev{\hs_{x_0}^x}_0 + \sum_{i} 2 i (1-g/g_c) \delta \ev{\hs_i^z} \ev{\hs_i^z \hs_{x_0}^{z} }_0 + \dots\), in which we only show the scaling part and ignore the finite terms. These terms can be dropped since they should satisfy the constraint equations for \(h=0\).
Note that \(\sum_i \ev{\hs_i^z \hs_{x_0}^{z}}_0 \simeq \int dx \ev{\hs^z(x) \hs^{z}(0) }_0 \propto \int dx |x|^{-1/4} \propto  \abs{1-g/g_c}^{3/4} \), and similarly \(\sum_i \ev{\hs_i^x \hs_{x_0}^{x} }_0 \propto  \abs{g-g_c}^{3/4} \). Also \(\ev{\hs_{x_0}^{z} }_0\) is finite and of \(\mathcal{O}(1)\). By requiring these terms to have the same scaling exponent, we find \(a = 1/8 + 3/4 + 1 = 15/8\), in agreement with the scaling limit results. Moreover, the VES results potentially can go beyond the scaling limit. We could consider \(\delta \ev{\hs^z} \propto |1-g/g_c|^{b}\), then our result indicates that \(b = a - 7/4\). VES is also applicable for arbitrary finite perturbation formula without assuming any scaling form for \(h\).

\emph{Application III: antiferromagnetic Heisenberg models.}---
In previous examples, we consider well controlled perturbation scenarios. We find that, even at a quantum critical point, the perturbation formula worked well. In this part, we briefly consider the Heisenberg models of quantum spin-1/2s and demonstrate a scenario where the variational approach is necessary. Here we lack a controlling parameter \(\lambda < 1\). But nonetheless we decompose the Heisenberg interaction into an Ising part and a transverse part: \(H_{\text{Heins.}} = H_{\text{Ising}} + H_{XX}\) with \(H_{\text{Ising}} = J \sum_{ij} \hs^z_i \hs^z_j\) and \(H_{\text{XX}} = J_{\perp} \sum_{ij} (\hs^x_i \hs^x_j + \hs^y_i \hs^y_j)\). We can start with one of the classical \(\text{N\'eel}\) state, or depending on the details, we could also start with a quantum superposition of the two-fold degenerate \(\text{N\'eel}\) states with variational parameters. For the Heisenberg limit  \(J_\perp/J \rightarrow 1^-\), it is reasonable to invoke the variational formula Eq. \eqref{eq:6}. We shall demonstrate that it is necessary.

For this problem, it is convenient to use the following COBS: \(\mathscr{U} = \{2 \hs^\alpha_i, 4 \hs^{\alpha_i}_i \hs^{\alpha_j}_j, \dots \}\).
The ground state of \(H_{\text{Ising}}\) is two-fold degenerate. Let \(\ket{\Psi_1} = \ket{\ua\da\ua\da \dots} ,
\ket{\Psi_2} = \ket{\da\ua\da\ua \dots} ,
\ket{\Psi_0} = \sin \theta \ket{\Psi_1} + \cos \theta \ket{\Psi_2}\). The important correlations of \(\ket{\Psi_0}\) are \(\ev{\hs^z_i - \hs^z_j}_0 = (-1)^{i-j} \cos 2\theta, \ev{\hs^z_i \hs^z_j}_{0,c} = (-1)^{i-j} \sin^2(2\theta) / 2, \ev{\hs^{x(y)}_i \hs^{x(y)}_j} = 0\).

First, let us try to apply the perturbation formula Eq. \eqref{eq:3} for a pure \(\text{N\'eel}\) state with \(\theta  = 0\). A suitable choice for \(\hu^{\alpha_0}\) is \(\hu^{\alpha_0} = 2 (\hs^{x}_{i_0} \hs^y_{i_0+1} - \hs^{y}_{i_0} \hs^x_{i_0+1})\). We obtain \(\ev{[H_{\text{XX}}, \hu^{\alpha_0}]} = -i J_\perp \ev{\hs^z_{i_0} - \hs^z_{i_0+1}}_0 = - i J_\perp (-1)^{i_0}\). For perturbation term, we have \(\sum_{\hu^\alpha} \ev{\{ [H_{\text{Ising}}, 4 \hs^{x}_{i_0} \hs^y_{i_0+1}], \hu^\alpha \}}_0 = 4 i J  \sum_{\hu^\alpha, \bf{-1 , +2}} \ev{\{ \hs^z_{i_0 - 1} \hs^y_{i_0} \hs^y_{i_0+1} -  \hs^z_{i_0 + 2} \hs^x_{i_0} \hs^x_{i_0+1}, \hu^\alpha \} }_0 \delta \ev{\hu^\alpha}_1\), where \(\sum_{\bf{-1 , +2}}\) denotes summation over all nearest neighbors of \(i_0\) and \(i_0+1\) respectively. We find that only one term \(\hu^\alpha = (\hs^x_{i_0} \hs^x_{i_0+1} + \hs^y_{i_0} \hs^y_{i_0+1}) = \hat{R}_{1}\) contributes, which is just the in-plane RVB correlation as expected. The corresponding variation of the state satisfies \(\frac{J_\perp}{J} + 4 (Z_{nn}-1)  \delta \ev{\hat{R}_{1}} \simeq 0\), where \(Z_{nn}\) denote the number of nearest neighbor bonds of \(H_0\). This bring the energy down to \((-2/3) J \simeq 0.6667 J\) per site when setting \(J_\perp \rightarrow J\) and \(Z_{nn} = 4\) for a 2D square lattice Heisenberg model, which agree with spin wave theory\cite{singh-1989-therm-param}.

However, this choice of \(\hu^{\alpha_0}\) breaks down if we start with \(\theta  = \pi/4\) which leads to \(\ev{\hs^z_{i_0} - \hs^z_{i_0+1}}_0 = 0\). Instead, we need to use \(\hu^{\beta_0} = 2 (\hs^z_{i_0-1} - \ev{\hs^z_{i_0-1}}_0) (\hs^{x}_{i_0} \hs^y_{i_0+1} - \hs^{y}_{i_0} \hs^x_{i_0+1})\), which leads to \(\ev{[H_{\text{XX}}, \hu^{\alpha_0}]}_0 = -i J_\perp \ev{\hs^z_{i_0-1} (\hs^z_{i_0} - \hs^z_{i_0+1})}_{0,c}\). More generally, to start with an arbitrary \(\theta\), we need both equations, which are generally linearly independent.

Moreover, while we see that the bare perturbation formula does yield RVB correlations as expected, it has no correction to \(\ev{\hs^z}\), which is an important benchmark. We also want to understand how RVB bonds on larger distances emerge.
To improve on that, we find it necessary to invoke the variational formula Eq. \eqref{eq:6} up to the second order. The first order VES generated by \(\hu^{\alpha_0}\) and \(\hu^{\beta_0}\) are equivalent.

With the help of QuantumAlgebra.jl\cite{QuantumAlgebra.jl,Sanchez-Barquilla2020}, we find the following VES
\begin{align}
  \begin{split}
   & \frac{J_\perp}{J} + \Delta_z + 4 (Z_{nn}-1) \delta \ev{\hat{R}_{1}} + 2 (Z_{nn}-1)^2 \\
   & \times \delta \ev{\hat{R}_{1}} \delta \ev{\hat{R}_{3}} + 4 (Z_{nn}-1)^2 \delta \ev{\hat{R}_{3}}^2 \simeq 0,
    \end{split} \label{eq:12}
    \\
    \begin{split}
  &  \frac{J_\perp}{J} + 4 (Z_{nn}-1) \delta \ev{\hat{R}_{1}} + 12 \delta m_z^2  - 2 (Z_{nn}-1)^2  \\
  & \times  \frac{m_z^2}{1-m_z^2} \delta \ev{\hat{R}_{1}} \delta \ev{\hat{R}_{3}} - 4 (Z_{nn}-1)^2 \frac{m_z^2}{1-m_z^2} \delta \ev{\hat{R}_{3}}^2 \simeq 0.
  \end{split}
\label{eq:13}
\end{align}
where \(\hat{R}_{3} = \hs^x_{i_0-1} \hs^x_{i_0+2} + \hs^y_{i_0-1} \hs^y_{i_0+2}\) is the next-next-nearest-neighbor in-plane RVB bond, \(m_z = \ev{ 2 \hs^z_{i_0}}_0\) is the normalized magnetization and its variation is \(\delta m_z = \delta \ev{2 \hs^z_{i_0} }\) and \(\Delta_z = \sum_i (-1)^{i+1} \delta \ev{2 \hs^z_i}/ m_z\). All terms involve the next-nearest-neighbor cancels when we assume translational invariance.

The most notable term is the \(\Delta_z\) term, which diverges if \(\delta \ev{2 \hs^z_i} \propto (-1)^{i+1}\) if one applies the translational invariance. Therefore, it seems to be forbidden to be finite, i.e. \(\delta \ev{2 \hs^z_i} = 0\). However, the appearance of this term is a consequence of spontaneous symmetry breaking(SSB) of a quantum Hamiltonian\cite{emch-1980-algeb-approac}. In fact, a SSB state density matrix does not satisfy the condition \([\rho, \mH] = 0\), which is not an issue for our starting point, the AFM Ising model. On the other hand, we know from spin-wave theories that reduction of the SSB order parameter, i.e. the staggered magnetization, does happen. Note that \(\Delta_z\) is precisely the SSB order parameter, we argue that the divergence of \(\Delta_z\) due to reduction of \(m_z\) should be canceled by SSB order's contribution to \([\rho, \mH]\).

Moreover, if we consider the thermodynamic limit, then \(\Delta_z = 0\) can also be satisfied by spatially fluctuating \(\delta \ev{2 \hs^z_i}\) over large distances. We consider that such variations correspond to the spin-wave excitations or magnons. The presence of such \(\delta \ev{2  \hs^z_i}\) is interpreted as a signature of spin-wave excited states.
Traditional spin-wave theory studies\cite{Dyson1956,singh-1989-therm-param} showed that interactions  between magnons due to \(H_{Ising}\) eventually led to reduction of \(m_z\) in the ground state.
Thus, we also expect that the VES determining \(\delta \ev{2  \hs^z_i}\) should come from higher order terms as well as coupling different branches of variations. How we can systematically achieve that will our future goal.

Next, we first ignore the terms involve \(\delta m_z\), Eq. \eqref{eq:12} and \eqref{eq:13}, although now linearly independent, allow for two solutions: \(\delta \ev{\hat{R}_{1}} = -1/12,~ \delta \ev{\hat{R}_{3}} = 1/24\) and a trivial one \(\delta \ev{\hat{R}_{1}} = -1/12,~ \delta \ev{\hat{R}_{3}} = 0\), identical to that of the first order VES. The ground state energy remains unchanged despite a longer RVB bond is found.

Now we take \(\delta m_z\) into consideration. In the second order terms, \(\delta m_z\) is cut off to local terms due to the linked cluster theorem. Since we know \(\delta \ev{2  \hs^z_i}\) is dominated by long-wave length fluctuations, locally we can still approximate it well by a constant, which we are computing here. Instead, we install a \(\delta m_z\) behavior by hand according to  spin-wave theories. We consider \(\delta m_z \simeq (- 2/9 - 8/225 m_z^2) m_z\) according to Ref. \cite{singh-1989-therm-param} which is for \(m_z = 1\). With that, we can minimize the ground state energy with respect to \(m_z\). For a 2D square lattice, we find, \(\ev{\hs^z_0} \simeq = m_z/2 \simeq \pm 0.366, ~ E \simeq - 0.675 J\). We consider it reasonably good given that the calculation involves correlation over merely four sites. Using \(\delta m_z\) at \(m_z = 1\) is likely a under-estimate. We also left out a variation of term of \((\delta\ev{\hs^z_i \hs^z_j})^2\) for simplicity, which should increase the ground state energy back up.

\emph{Discussion and Conclusion}.---
With a rigorous and generic language, a COBS representation of density matrices and their algebraic properties, we derived a set of variational equations of states, coupled algebraic equations of variations of correlation functions, for interacting quantum Hamiltonians. While the raw VES are complicated for many-body systems on lattices, we find that a linked cluster theorem can greatly simply them. Then we successfully applied the VES approach to three different problems and obtained results comparable established ones with less efforts. The VES formalism is generally applicable to any problems on lattices in any dimensions, including fermionic problems, hybridized problems such as the Kondo lattice problems.

In the current work, we restricted to algebras and COBS in real spaces for bosons. It is generally considered more cumbersome to work with momentum space algebras for noncanonical operators due to the nontrivial momentum dependence of the algebra. For VES, it could be more convenient in the thermodynamic limit, which turns the algebraic equations into integral equations in momentum space.
VES can also be generalized to finite temperatures. Ref. \cite{fano-1957-descr-states} discussed such conditions for quantum mechanical problems.

The VES alone already provide a highly efficient way to compute both energy and correlation functions simultaneously. It is natural and possible to combine its analytic power with other variational numerical approaches. For example, the recently proposed bootstrap approach for quantum mechanical and many-body systems\cite{han-2020-quant-many,han-2020-boots-matrix}. Moreover, the many-body correlation functions naturally connect with the intrinsic structures of tensors in tensor-network methods. If such connection can be established explicitly, the VES could help improve tensor methods or may be directly integrated with them.

\emph{Acknowledgement.}---We thank  Rong Yu, Yang Qi, Jianda Wu, Zhentao Wang and Zhiyuan Xie for insightful discussion. The work at Anhui University was supported by the National Key R\&D Program of the MOST of China (Grant No. 2022YFA1602603) and the Startup Grant number S020118002/002 of Anhui University.

%\bibliography{../library,../../org/notes,../../org/research/refs}


\end{document}